\documentclass[fleqn,twoside,twocolumn,nofootinbib]{revtex4} 
\usepackage[web]{ujp} 
\begin{document}
\title[GENERALIZED STILLINGER--DAVID POTENTIAL]
{GENERALIZED STILLINGER--DAVID POTENTIAL}
\author{I.V. ZHYGANIUK}
\affiliation{I.I. Mechnikov Odesa National University}
\address{2, Dvoryans'ka Str., Odesa 65026, Ukraine}
\email{ivz@ukrpost.ua}


\setcounter{page}{225}%
\maketitle

\begin{abstract}
We present an improved version of the Stillinger--David polarization potential of the intermolecular interaction in water. A clear algorithm of construction of a function describing the oxygen-hydrogen interaction in water molecules is formulated. A new approach to the modeling of a function screening the charge-dipole interaction on small distances is developed. To describe the long-range asymptotics of the intermolecular potential, the bare Stillin\-ger--David potential is supplemented by a term related to the interaction of dipole moments of oxygen ions. In addition, we introduce a term involving a deformation of the electron shells of oxygen ions to the polarization component. These corrections allow us to successfully reproduce all essential results of quantum mechanical calculations of the interaction energy for water molecules obtained by Clementi. Analyzing the behavior of the dipole moment of a water molecule as a function of the intermolecular distance, we obtain the estimate of irreducible two-particle effects in water.
\end{abstract}

\section{Introduction}

Astonishing properties of water are directly related to the formation of hydrogen bonds between water molecules \cite{BulavinKarm,BulavinFisen,AdamenkoBulavin,AntonchenkoDavydov}. Various potentials are used to simulate the influence of hydrogen bonds, as well as electrostatic multipole interactions between water molecules \cite{AntonchenkoDavydov,AntonchenkoIlyin,Antonchenko,Bernal,RahmanStillinger,Jorgensen,Kistenmacher,Poltev,StillingerDavid}. One has to distinguish between rather simplified potentials, which mainly describe the interaction between water molecules \cite{Bernal,RahmanStillinger,Jorgensen,Kistenmacher} , and potentials, which describe both the molecule-to-molecule interaction and the interaction between model charges inside a water molecule \cite{AntonchenkoDavydov,AntonchenkoIlyin,Antonchenko,StillingerDavid}. As a result, the calculation, e.g., of vibrational spectra of water molecules becomes possible.

In the framework of the Malenkov--Grokhlina--Poltev potential model~\cite{Poltev}, a water molecule is simulated as that composed of three effective charges: one positive and two negative charges located at the centers of the oxygen and two hydrogen atoms, respectively. The Jorgensen potential  \cite{Jorgensen} also represents a water molecule as three effective charges: the positive charges are centered at hydrogen atoms, but the negative charge is displaced with respect to the oxygen anion center. It was done to match the dipole moment of a water molecule. The model of intermolecular interaction potential turned out optimal, provided that the charge values are fractional. Note that the Jorgensen and Malenkov--Grokhlina--Poltev potentials, owing to their simple structure, are widely used, while simulating the behavior of aqueous systems by methods of molecular dynamics.

The Stillinger-David (SD) potential~\cite{StillingerDavid, StillDav} is one of the most fruitful models of interaction between particles, which was used to describe the formation of hydrogen bonds in water. In the framework of this potential, a water molecule is represented by three charges, and the oxygen anion is supposed to be polarizable.  With the help of the SD potential, both the intermolecular interaction and the vibrational spectra of water molecules are reproduced satisfactorily~\cite{AntonchenkoDavydov, StillingerDavid, StillDav}. The approach by Stillinger and David allows one to obtain a good agreement with experiment for the dipole moment of a water molecule and the angular dependence of the interaction between molecules which corresponds to the formation of hydrogen bonds. This potential was also used, while constructing the autocorrelation functions for translational and angular velocities of water molecules, as well as when calculating their self-diffusion coefficient~\cite{AntonchenkoDavydov}. The SD potential was also used in works~\cite{Lokotosh,Zabrodskii,LokotoshMalomuzh,ZabrodskiiLokotosh}, where the spectra of acoustical and optical excitations in crystalline ice and the ice dielectric permittivity were calculated~\cite{ZabrodskiiLokotosh}.

In works~\cite{AntonchenkoDavydov,AntonchenkoIlyin,Antonchenko}, a modification of the Stillinger--David potential--the modified polarization model (MPM) -- was proposed. This model substantially simplifies the form of potentials for the oxygen--hydrogen and oxygen--oxygen interactions and uses only one screening function, $S(r)$, rather than two, $1-L(r)$ and $1-K(r)$, as the SD potential does. As a result, the values of force constants were calculated more precisely, and the computation procedure was made simpler~\cite{AntonchenkoDavydov}.

At the same time, the results obtained by Stillinger and David, as well as by the authors of MPM do not agree with the results of quantum-chemical calculations carried out by Clementi~\cite{Kistenmacher} at distances shorter than about 2.5~\AA. A comparison of MPM and SD potentials with the interaction energy that was determined in work~\cite{Kistenmacher} evidences their unsatisfactory agreement at distances that either correspond to or are much longer than those corresponding to the dimer formation.
	
In this work, a generalized version of the Stillinger-David potential (GSD) has been developed. In this model, we try to combine the most successful features of the SD and MPM potentials. In particular, (i) the functions that describe the energies of the oxygen-hydrogen and oxygen-oxygen interactions are similar to the expressions obtained in the MPM; (ii) in order to determine the function that describes the oxygen-hydrogen interaction energy in a water molecule, a very rigorous algorithm has been formulated. It takes into account the fact that the solutions of the system of algebraic equations for the coefficients of a polynomial that is used to approximate the screening function $1-L(r)$ have a very irregular character; (iii) besides the Coulomb and polarization components of the SD potential, the interaction between the induced dipole moments of oxygen atoms is taken into consideration, which is very important for a correct reproduction of the dipole-dipole interaction energy between two water molecules separated by long enough distances. As shown below, the screening function $1-L(r)$ changes monotonously, thus being substantially different from that given in works~\cite{StillingerDavid, StillDav}. Owing to the aforesaid improvements, the results of quantum-chemical calculations by Clementi were reproduced quite successfully.

In Section 2 of the work, a modified version of the Stillinger-David (MSD) potential is presented. It is used to obtain the dipole-dipole asymptotics of the interaction energy between water molecules. In Section 3, the laws of interaction between the hydrogen atoms and the oxygen one, which are included into a water molecule, are generalized. A short discussion of the results obtained is presented at the end of the article.

\section{Modified Stillinger-David Potential}

In the framework of the Stillinger-David potential approximation~\cite{StillingerDavid}, the energy of interaction between two water molecules is expressed by the formula
\begin{equation}
\Phi  = \Phi _{\rm I}  + \Phi _{{\rm I}{\rm I}}  + \Phi _{{\rm I}{\rm I}{\rm I}}.
\end{equation}

The first contribution $\Phi _{\rm I}$ in Eq. (1) is associated with a direct Coulomb interaction between oxygen and hydrogen atoms in a water molecule,
\begin{equation}
\Phi _{\rm I}  = \sum\limits_{i,j = 0}^{i,j = 2} {\frac{{q_i q_j }}{{r_{i j} }}},
\end{equation}
where the subscripts $i, j = 0, 1, 2$. The subscript i enumerates the charges of one water molecule ($i = 0$ corresponds to the oxygen charge, whereas $i = 1, 2$ to the hydrogen ones), and j enumerates the charges of the other molecule. The charges are measured in terms of elementary charge units, so that $q_0 = - 2$ and $q_1 = q_2 = 1$. The Coulomb interaction between hydrogen and oxygen charges at large distances is reduced to the interaction between the dipole moments ${\boldsymbol {\mu }}_{\scriptscriptstyle {\mathrm H}}$ of water molecules associated with the charges of hydrogen atoms in those molecules.

The second contribution $\Phi_{{\rm I}{\rm I}}$ corresponds to the repulsion potential of a hydrogen atom from the electron shell of the oxygen atom. It is approximated by the Born exponential dependence
\[
\Phi _{{\rm I}{\rm I}}  = b_1 \left[ {\sum\limits_{i = 1,2}^{} {\frac{{e^{ - \rho _1 r_{i {\scriptscriptstyle{\mathrm O}}_2 } } }}{{r_{i {\scriptscriptstyle{\mathrm O}}_2 } }}}  + \sum\limits_{j = 1,2}^{} {\frac{{e^{ - \rho _1 r_{{\scriptscriptstyle{\mathrm O}}_1 j} } }}{{r_{{\scriptscriptstyle{\mathrm O}}_1 j} }}} } \right]+
\]
\begin{equation}
+ \frac{{b_2 e^{ - \rho _2 r_{{\scriptscriptstyle{\mathrm O}}_1 {\scriptscriptstyle{\mathrm O}}_2 } } }}{{r_{{\scriptscriptstyle{\mathrm O}}_1 {\scriptscriptstyle{\mathrm O}}_2 } }},
\end{equation}
where $b_1$ and $b_2$ are the amplitudes of hydrogen-oxygen and oxygen-oxygen, respectively, repulsion energies; $\rho _1$ is the reciprocal range of action of repulsion forces between the hydrogen atoms and the electron shell of oxygen; and $\rho _2$ is the reciprocal range of action of the repulsion forces between the electron shells of oxygen atoms.

The third contribution $\Phi _{{\rm I}{\rm I}{\rm I}}$ in Eq. (1) is responsible for the potential of interaction between the point-like charges of the first molecule with the polarizable oxygen atom of the second one. The oxygen atom is polarized under the influence of a field created by the charges of hydrogen atoms belonging to the same water molecule, as well as by the charges of hydrogen atoms and the polarized oxygen atom belonging to the second molecule. The oxygen polarization gives rise to an emergence of the dipole moment ${\boldsymbol {\mu }}_{\scriptscriptstyle{\mathrm O}}$, which characterizes the deformation of oxygen electron shells. Hence, the polarization contribution $\Phi _{{\rm I}{\rm I}{\rm I}}$ can be written down as follows:
\[
\Phi _{{\rm I}{\rm I}{\rm I}}  =  \frac{{({\boldsymbol {\mu }}_{{\scriptscriptstyle{\mathrm O}}_1 }  \cdot  {\bf{r}}_{{\scriptscriptstyle{\mathrm O}}_1 {\scriptscriptstyle{\mathrm O}}_2 }  ) q_{{\scriptscriptstyle{\mathrm O}}_2 } }}{{r_{{\scriptscriptstyle{\mathrm O}}_1 {\scriptscriptstyle{\mathrm O}}_2 }^3 }}[1 - L(r_{{\scriptscriptstyle{\mathrm O}}_1 {\scriptscriptstyle{\mathrm O}}_2 } )] +
\]
\[
 +  \frac{{({\boldsymbol {\mu }}_{{\scriptscriptstyle{\mathrm O}}_2 }  \cdot  {\bf{r}}_{{\scriptscriptstyle{\mathrm O}}_1 o_2  }  )q_{{\scriptscriptstyle{\mathrm O}}_1 } }}{{r_{{\scriptscriptstyle{\mathrm O}}_1 {\scriptscriptstyle{\mathrm O}}_2 }^3 }}[1 - L(r_{{\scriptscriptstyle{\mathrm O}}_1 {\scriptscriptstyle{\mathrm O}}_2  } )] +
\]
\[
 + \left[ {\sum\limits_{j = 1,2}^{} {\frac{{({\boldsymbol {\mu }}_{{\scriptscriptstyle{\mathrm O}}_1 }  \cdot  {\bf{r}}_{{\scriptscriptstyle{\mathrm O}}_1 j}  )q_j }}{{r_{{\scriptscriptstyle{\mathrm O}}_1 j}^3 }}[1 - L(r_{{\scriptscriptstyle{\mathrm O}}_1 j} )]}  + } \right.
\]
\begin{equation}
 + \left. {\sum\limits_{i = 1,2}^{} {\frac{{({\boldsymbol {\mu }}_{{\scriptscriptstyle{\mathrm O}}_2 }  \cdot  {\bf{r}}_{{\scriptscriptstyle{\mathrm O}}_2 i }  )q_i }}{{r_{{\scriptscriptstyle{\mathrm O}}_2 i}^3 }}[1 - L(r_{{\scriptscriptstyle{\mathrm O}}_2 i } )]} } \right],
\end{equation}
where  ${\boldsymbol {\mu }}_{{\scriptscriptstyle{\mathrm O}}_1 }$, and ${\boldsymbol {\mu }}_{{\scriptscriptstyle{\mathrm O}}_2 }$ are the dipole moments of oxygen atoms of the first and the second, respectively, water molecule; and $1 - L(r)$ is the screening function (see below). The dipole moments are determined in the molecular coordinate system (MCS), the origin of which coincides with the center of mass of the oxygen atom in the water molecule. At large distances, the polarization contribution $\Phi_{{\rm I}{\rm I}{\rm I}}$ is reduced to the interaction between the polarized oxygen atom of the first molecule and the charges of the second molecule.

One can verify that, at distances that exceed the dimensions of water molecules very much, the interaction potential $\Phi$ looks like
\begin{equation}
\Phi  = \Phi _d ({\boldsymbol {\mu }}_{\scriptscriptstyle{\mathrm{1H}}} ,{\boldsymbol {\mu }} _{\scriptscriptstyle{\mathrm{2H}}} ) + \Phi _d ({\boldsymbol {\mu }} _{\scriptscriptstyle{\mathrm{1H}}} ,{\boldsymbol {\mu }} _{{\scriptscriptstyle{\mathrm O}}_2 } ) + \Phi _d ({\boldsymbol {\mu }} _{\scriptscriptstyle{\mathrm{2H}}} ,{\boldsymbol {\mu }} _{{\scriptscriptstyle{\mathrm O}}_1 } )\, ,
\end{equation}
where
\begin{equation}
\Phi_d ({\boldsymbol {\mu }}_1 ,{\boldsymbol {\mu }}_2 ) = \frac{{\left( {{\boldsymbol {\mu }}_1  \cdot {\boldsymbol {\mu }}_2 } \right)}}{{r_{{\scriptscriptstyle{\mathrm O}}_1 {\scriptscriptstyle{\mathrm O}}_2 }^3 }} - \frac{{3\left( {{\boldsymbol {\mu }}_1  \cdot {\boldsymbol {r}}_{{\scriptscriptstyle{\mathrm O}}_1 {\scriptscriptstyle{\mathrm O}}_2 } } \right)   \left( {{\boldsymbol {\mu }}_2  \cdot {\bf{r}}_{{\scriptscriptstyle{\mathrm O}}_2 {\scriptscriptstyle{\mathrm O}}_1 } } \right)}}{{r_{{\scriptscriptstyle{\mathrm O}}_1 {\scriptscriptstyle{\mathrm O}}_2 }^5 }} . \, \, \, \,
\end{equation}
In formula (5), ${\boldsymbol {\mu }}_{\scriptscriptstyle{\mathrm{1H}}}$ and ${\boldsymbol {\mu }}_{\scriptscriptstyle{\mathrm{2H}}}$ are the contributions to the dipole moment of water molecules resulting from the spatial arrangement of hydrogen charges. The oxygens are at the origin of the MCS, and their contributions ${\boldsymbol {\mu }} _{{\scriptscriptstyle{\mathrm O}}_1 }$ and ${\boldsymbol {\mu }} _{{\scriptscriptstyle{\mathrm O}}_2 }$ to the dipole moment may arise only due to their polarization by the electric field of hydrogen atoms. Asymptotics (5) is incorrect, because the energy of interaction between water molecules at large enough distances is determined by their total dipole moments,
\begin{equation}
{\boldsymbol {\mu }} = {\boldsymbol {\mu }}_{\scriptscriptstyle{\mathrm H}}  + {\boldsymbol {\mu }}_{\scriptscriptstyle{\mathrm O}} .
\end{equation}

We note that the influence of hydrogen atoms that belonging to the same molecule is by two orders of magnitude larger than the influence of charges in the second molecule. This fact completely agrees with the order of magnitude of a relative variation in the frequency of valence vibrations in the water molecule, when changing from the liquid-water to the saturated-vapor state~\cite{Eisenberg}.

Unfortunately, asymptotics (5) is also characteristic for the modified polarization model (MPM) that was developed in works~\cite{AntonchenkoDavydov,AntonchenkoIlyin,Antonchenko}.

To correct the asymptotics of formula (1), we have to pass to the modified Stillinger-David potential,
\begin{equation}
\Phi  = \Phi _{\rm{I}}  + \Phi _{{\rm{II}}}  + \Phi _{{\rm{III}}}  + \Phi _{{\rm{IV}}},
\end{equation}
\begin{figure}[!h]
\centering
\includegraphics[natwidth=1680, natheight=1976, scale=0.14]{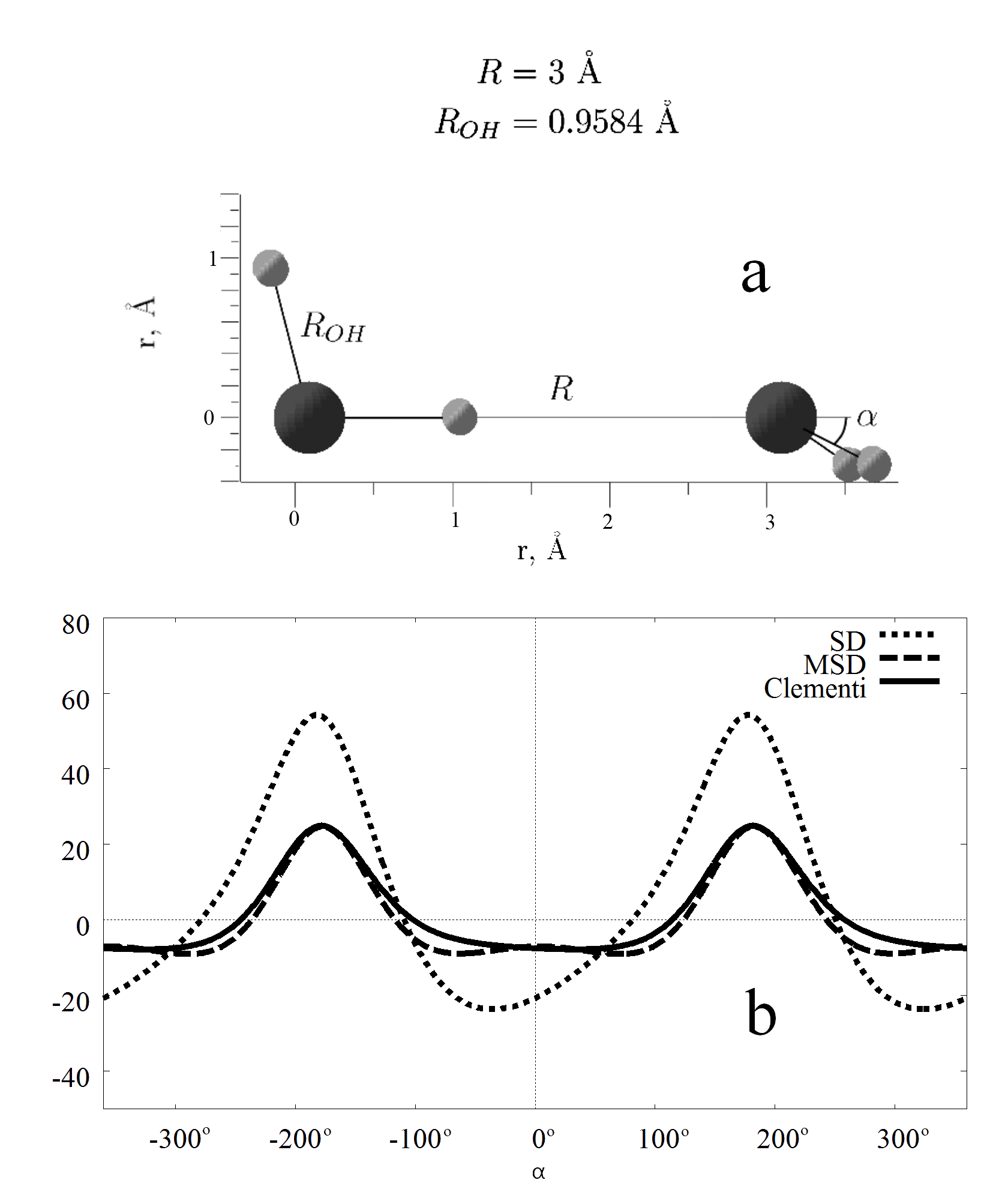}
\vskip-3mm\caption[Configuration of molecules]{(a) Configuration of molecules and (b) the angular dependence (at $\alpha = - {360}^{\circ} \div {360}^{\circ}$) of the interaction energy between water molecules calculated with the use of the SD (dotted curve) and MSD (dashed curve) potentials and the Clementi function (solid curve)}\label{fig1}
\end{figure}
where the component $\Phi _{{\rm{IV}}}$ looks like
\begin{equation}
\Phi _{{\rm{IV}}}  = \Phi _d ({\boldsymbol {\mu }}_{ {\scriptscriptstyle{\mathrm O}}_1 } ,{\boldsymbol {\mu }}_{{\scriptscriptstyle{\mathrm O}}_2 } )[1 - K({{r_{{{\scriptscriptstyle{\mathrm O}}_1 } {{\scriptscriptstyle{\mathrm O}}_2 } } } \mathord{\left/
 {\vphantom {{r_{{{\scriptscriptstyle{\mathrm O}}_1 } {{\scriptscriptstyle{\mathrm O}}_2 } } } a}} \right.
 \kern-\nulldelimiterspace} a})] .
\end{equation}
Here, ${1 \mathord{\left/
 {\vphantom {1 a}} \right.
 \kern-\nulldelimiterspace} a}$ is a coefficient that modifies the screening function $1 - K(r)$ for the interaction between the dipole moments of oxygen atoms in two neighbor water molecules. The optimal fitting of Clementi's data (see Fig.~\ref{fig1}) was obtained at   $a = 2.235$.

It should be noted that the component $\Phi _{{\rm{IV}}}$~(Eq.(9)) of the MSD potential supplements the SD potential by including the dipole-dipole interaction between polarized oxygens in water molecules. The effect of the dipole moment screening for oxygen atoms in water molecules is described by the function $1 - K({r \mathord{\left/
 {\vphantom {r a}} \right.
 \kern-\nulldelimiterspace} a})$.
The screening function $1 - K({r \mathord{\left/
 {\vphantom {r a}} \right.
 \kern-\nulldelimiterspace} a})$ is similar to the function $1 - K(r)$ which was used in works~\cite{StillingerDavid, StillDav} to describe the dipolecharge interaction in the SD potential (see below).

Owing to the screening functions $1 - K({r \mathord{\left/
 {\vphantom {r a}} \right.
 \kern-\nulldelimiterspace} a})$, the interaction energy between water molecules almost does not change at short distances. Meanwhile, at large disances, formula (8) gives the correct asymptotics,
\begin{equation}
\Phi  \to \Phi _d ({\boldsymbol {\mu }}_1 ,{\boldsymbol {\mu }}_2 ).
\end{equation}

The Stillinger-David potential and its modified version are compared in Fig. 1,b.
The configuration of water molecules is depicted in Fig. 1,a.

\section{Generalized Stillinger--David Potential}

In this Section, the modified Stillinger--David potential is generalized further. The structure of the interaction potential between two water molecules (8) is adopted to remain invariant. Only the character of the interaction between oxygen atoms and hydrogen atoms in both the molecule and its neighbor changes considerably. In particular, the form of the screening function $1-L(r)$ is improved, and the fact is taken into account that the polarization of oxygen atoms depends, to some extent, on the electric field that emerges owing to a deformation of the electron shells of oxygen atoms induced by their immediate contact.

\subsection{Behavior of screening functions}

In the Stillinger--David approach, two screening functions, $1-K(r)$ and $1-L(r)$, are used. We adopt that the function $1-K(r)$ is equal to that obtained in work~\cite{StillingerDavid}:
\begin{equation}
1 - K(r) = {{r^3 } \mathord{\left/
 {\vphantom {{r^3 } {[r^3  + F(r)]}}} \right.
 \kern-\nulldelimiterspace} {[r^3  + F(r)]}},
\end{equation}
where
\begin{equation*}
 F(r) = 1.855785223(r - r_{{\scriptscriptstyle{\mathrm{OH}}}} )^2 \exp[{ - 8(r - r_{\scriptscriptstyle{\mathrm{OH}}} )^2 }] +
\end{equation*}
\begin{equation*}
+ 16.95145727 \exp[{ - 2.702563425 r}],
\end{equation*}
and $r_{\scriptscriptstyle{\mathrm{OH}}}  = 0.9584$ \AA.

The structure of the function $1-L(r)$ is approximated, similarly to what was done in work~\cite{StillingerDavid}, by a combination of an exponential function and a polynomial,
\begin{equation}
1 - L(r) = 1 - e^{ - L_0 r} (1 + L_1   r + L_2   r^2  + L_3   r^3  + L_4   r^4 ) .
\end{equation}
However, the relevant coefficients are determined now, by using a new rigorous algorithm. It should be noted that nine parameters are to be determined simultaneously; they include four coefficients $L_1$, $L_2$, $L_3$, and $L_4$ in expression (12) and four parameters $b_1$, $\rho_1$, $b_2$,  and $\rho_2$ in formula (3). The listed parameters were determined under the following conditions:
(i) the equilibrium distance between the oxygen and hydrogen atoms in a water molecule was accepted to be $r_{\scriptscriptstyle{\mathrm{OH}}}  = 0.9584$~\AA~\cite{AntonchenkoDavydov};
(ii) the angle between the directions from the oxygen atom toward hydrogen ones was taken $\theta  = 104.45^{\circ}$ in the equilibrium configuration (see work~\cite{AntonchenkoDavydov});
(iii) the force constants $\frac{{\partial ^2 \Phi }}{{\partial r_1^2 }} = 2064.114$ and $\frac{{\partial ^2 \Phi }}{{\partial r_1 \partial \theta }} = 91.5562$ were taken from the molecular spectral characteristics (see work~\cite{Smith});
(iv) the final result had to reproduce the energy of interaction between two water molecules that was obtained as a result of quantum-chemical calculations in work by Clementi and coauthors \cite{Kistenmacher}. We demanded that the function $1-L(r)$ should change monotonously, because its nonmonotonicity would undesirably affect the behavior of the derivatives of the interaction energy of a water molecule.

In addition, we adopted that the coefficient $L_0  - L_1$ in the linear term $r$ in the expansion of the function $1 - L(r)$ in a power series of $r$ in a vicinity of the point $0$ was equal to zero, i.e. $L_0 = L_1$. The coefficients $L_1$, $L_2$, $L_3$ and $L_4$  were fitted to reproduce the value of force constant
\begin{equation}
\frac{{\partial ^2 \Phi }}{{\partial r_1 \partial \theta }} = 91.5562
\end{equation}
and to satisfy the condition
\begin{equation}
\frac{{\partial \Phi }}{{\partial \theta }} = 0
\end{equation}
for the water molecule in the equilibrium configuration. Additionally, a requirement that the squared norm $\left\| {L_i } \right\|^2  = L_0 ^2  + L_1 ^2  + L_2 ^2  + L_3 ^2  + L_4 ^2$ should be minimal was introduced into the determination procedure of the coefficients $L_1$, $L_2$, $L_3$, and $L_4$.
A similar {\it a priori} condition was proposed by A.N. Tikhonov in work~\cite{Tikhonov} to distinguish stable normal solutions of the systems of linear equations. By definition, normal are those solutions, the moduli of which are close to zero.

In such a way, we obtained that the coefficients $L_0$, $L_1$, $L_2$, $L_3$, and $L_4$ are
\[
L_0  = 2.98, L_1  = 2.98,
\]
\begin{equation}
 L_2  = 0.92, L_3  = 4.7044, L_4  = 2.3580.
\end{equation}
In what follows, the magnitudes of physical quantities are given to an accuracy of 5 significant figures, because it is to this accuracy that the experimental values of force constants were determined.

\begin{figure}[!t]
\includegraphics[natwidth=1680, natheight=979, scale=0.145]{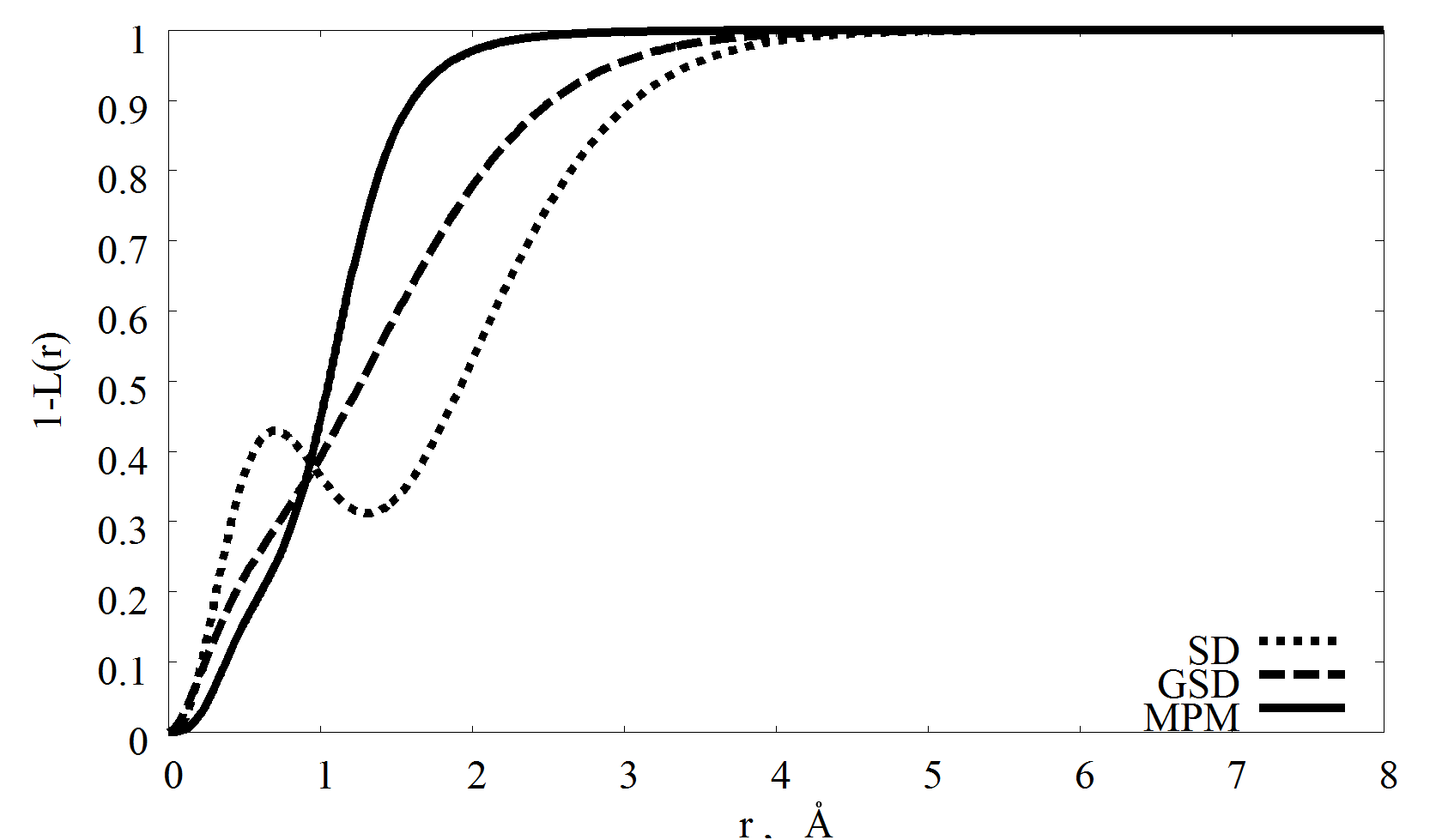}
\vskip-4mm\caption{Behavior of the screening functions in the SD, GSD, and MPM potentials}\label{fig2}
\end{figure}

The comparative behavior of the screening function ${1-L(r)}$ in the SD and GSD potentials and the corresponding function $S(r)$ in the MPM potential is presented in Fig.~\ref{fig2}. We emphasize that the screening function $1-L(r)$ is monotonous in the GSD potential (see
formula (15)), in contrast to its behavior in the SD potential, and more adequately reproduces the extension of screening region in comparison with the function $S(r)$ in the MPM potential~\cite{AntonchenkoDavydov}.

Note that the surface $F(L_0, L_1, L_2, L_3, L_4)$ is very ir-regular in the multidimensional space of parameters. As a consequence, the equation
$\frac{{\partial ^2 \Phi }}{{\partial r_1 \partial r_2 }} =  - 23.133$,  which should have been used for the determination of the
screening function coefficients $L_0, L_1, L_2, L_3$,  and $L_4$,
turns out incompatible with Eqs. (13) and (14). Coefficients (15) give the following value for this derivative:
$\frac{{\partial ^2 \Phi }}{{\partial r_1 \partial r_2 }} = 205.9$.
For a similar reason, the experimental
value of the derivative $\frac{{\partial ^2 \Phi }}{{\partial r_1 \partial r_2 }}$ was managed to be reproduced in neither the SD potential nor the MPM one.

For the determination of the parameters $b_1$  and $\rho _1$,
the following initial data were used: the aforesaid values
(see Eq. (15)) for the coefficients $L_0, L_1, L_2, L_3$, and $L_4$
of the screening function, the force constant
\begin{equation}
\dfrac{{\partial ^2 \Phi }}{{\partial r_1^2 }} = 2064.114,
\end{equation}
and the derivative
\begin{equation}
\dfrac{{\partial \Phi }}{{\partial r_1 }} = 0,
\end{equation}
\vskip-3mm
\begin{table}[!h]
\noindent\caption{Parameters of the MPM, SD, and GSD approximations}\vskip2.2mm\tabcolsep9.7pt
\noindent{\footnotesize \begin{tabular}{c c c c c c c c c c c}
 \hline%
 \multicolumn{1}{c}{   }
 & \multicolumn{1}{|c}{$L_0$}
 & \multicolumn{1}{|c}{$L_1$}
 & \multicolumn{1}{|c}{$L_2$}
 & \multicolumn{1}{|c}{$L_3$}
 & \multicolumn{1}{|c}{$L_4$} \\
 \hline
 MPM & - & - & - & - & -  \\
 SD  & 3.169 & 3.169 & 5.024 & -17.99 & 23.923  \\
 GSD & 2.98 & 2.98 & 0.92 & 4.704 & 2.358  \\
\hline
\end{tabular}
}
\vskip0.1mm
\noindent{\footnotesize \begin{tabular}{c c c c c}
\hline%
\multicolumn{1}{c}{ \qquad  \,  \, \, \,}
 & \multicolumn{1}{|c}{  \: $b_1$  \qquad}
 & \multicolumn{1}{|c}{  \:  $\rho _1$  \qquad}
 & \multicolumn{1}{|c}{ \:  $b_2$  \qquad}
 & \multicolumn{1}{|c}{  \: \, $\rho _2$  \, \,   \,  } \\
 \hline
 MPM  &  30335.16 & 5.678 & 3.5756 & 5.05 \\
 SD   & - & - & - & - \\
 GSD  & 3172.8 & 2.569 & 42129.1 & 2.59 \\
\hline
\end{tabular}
}
\end{table}
\begin{figure}[!t]
\includegraphics[natwidth=1680, natheight=979, scale=0.145]{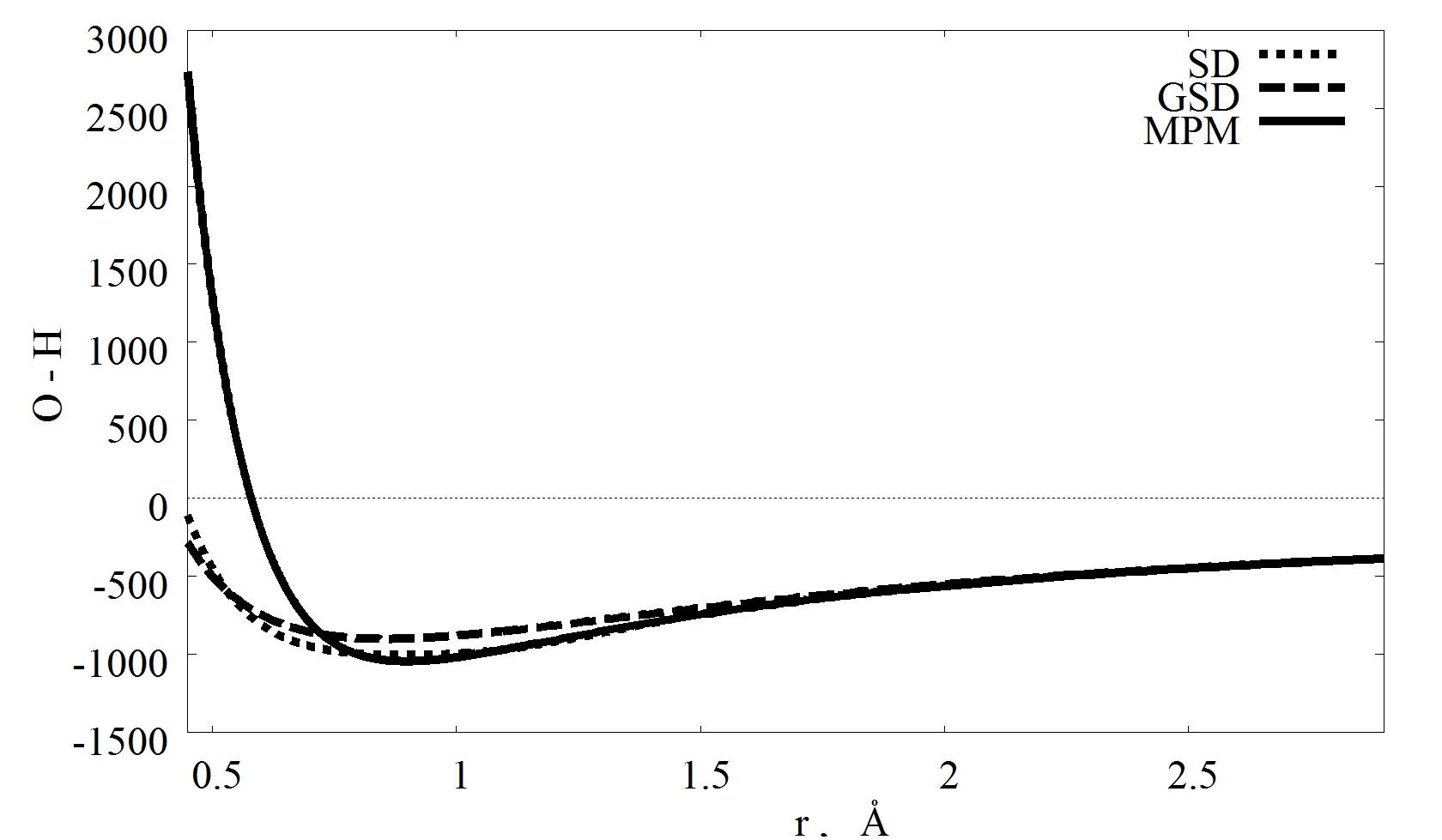}
\vskip-4mm\caption{Oxygen–hydrogen interaction energy in the SD, GSD, and
MPM potentials }\label{fig3}
\end{figure}
\vskip-2mm
\noindent{which correspond to the equilibrium configuration of a
water molecule. The corresponding values obtained for
the parameters of the GSD potential are}
\begin{equation}
b_1 = 3172.8, \rho _1 = 2.569.
\end{equation}
\vskip-8mm

\subsection{Parameters of the GSD Potential}

In Table 1, the values for all parameters in the GSD
potential are quoted, and a comparison with the corresponding values in the SD and MPM potentials is made.

The values of force constants in the GSD potential,
which were calculated for a water molecule with the help
of the parameters presented in Table 1 are listed in Table 2. For the sake of comparison, Table 2 also contains
the values for the same constants in the SD and MPM
potentials, as well as their values determined experimentally. One can see that the values of force constants obtained in the GSD potential are more exact than those
in the SD potential. Concerning the MPM potential,
the corresponding force constants are almost identical
to their experimental values.

A considerable advantage of the GSD potential with
respect to the SD and MPM ones becomes evident, when
calculating the interaction energy between two water
molecules (see below).

A comparison of functions that describe the oxygen--hydrogen interaction within the MPM, SD, and GSD
\vskip1mm
\begin{table}[!h]
\noindent\caption{Force constants for water molecule: experimental, MPM, SD, GSD }\vskip3mm\tabcolsep5.1pt
\noindent{\footnotesize \begin{tabular}{c c c c c}
 \hline%
 \multicolumn{1}{c}{\rule{0pt}{9pt} {\kern 7pt}    }%
 & \multicolumn{1}{|c}{   $\frac{{\partial ^2 \Phi }}{{\partial r_1^2 }}$ }
 & \multicolumn{1}{|c}{  $\frac{{\partial ^2 \Phi }}{{\partial \theta ^2 }}$ }
 & \multicolumn{1}{|c}{ $\frac{{\partial ^2 \Phi }}{{\partial r_1 \partial \theta }}$  }
 & \multicolumn{1}{|c}{  $\frac{{\partial ^2 \Phi }}{{\partial r_1 \partial r_2 }}$   } \\
 \hline
 Experimental \cite{Smith} & 2064.114 & 175.158 & 91.556 & -23.133 \\
 MPM  &  2064.114 & 175.158 & 91.556 & 286.94 \\
 SD   & 2064.114 & 167.342 & -34.485 & 117.57 \\
 GSD  & 2064.114 & 167.342 & 91.556 & 205.9 \\
 \hline
\end{tabular}
}
\end{table}

\eject

\begin{figure}[t]
\includegraphics[natwidth=690, natheight=524, scale=0.29]{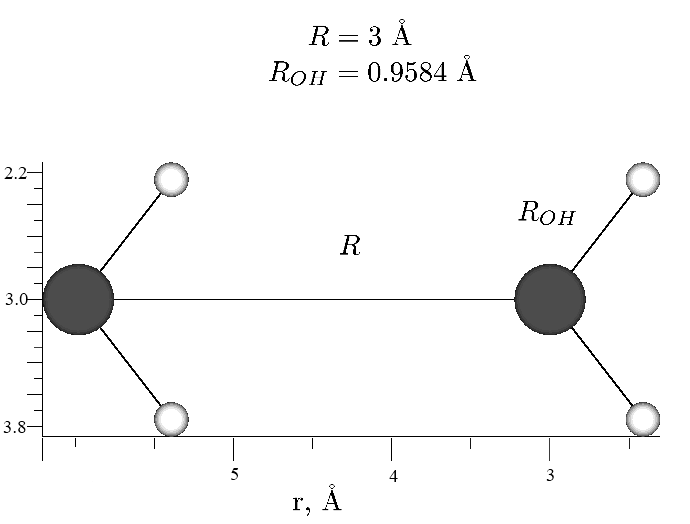}
\vskip-3mm\caption{Relative arrangement of two water molecules used in work \cite{Kistenmacher}
}
\end{figure}
\vskip-1mm
\noindent{potentials is made in Fig. 3. Figure 3 testifies that the
core of an oxygen atom is too rigid in the MPM. This circumstance is very important, when calculating the interaction energy between two water molecules at distances
that correspond to the dimer formation. }

\subsection{General structure of the GSD potential }

The structure of the interaction potential between two
water molecules in the GSD and MSD approximations
is the same (see Eq. (8)). However, the calculation procedures for the dipole moment of oxygen in a water molecule and the parameters $b_2$ and $\rho _2$, which describe the influence of the neighbor water molecule on the oxygen dipole moment, change. The circumstance is taken
into account that the electric field acting on the oxygen atom is a sum of the fields formed by the hydrogen atoms in the water molecule and the hydrogen and oxygen atoms belonging to the neighbor water molecules, on
the one hand, and a component that emerges as a result
of a deformation of the electron shells of oxygens, on the
other hand. The latter effect is particularly important
at distances between water molecules that correspond to
the dimer formation.

In accordance with all that, the dipole moment of an
oxygen atom is determined by the formula
\begin{equation}
{\boldsymbol {\mu }}_{ {\scriptscriptstyle{\mathrm O}}_1 } = \, -  \alpha \sum\limits_{} {\frac{{{\bf{\hat T}}_{{ {\scriptscriptstyle{\mathrm O}}_1 } { {\scriptscriptstyle{\mathrm O}}_2 } }  \cdot  {\boldsymbol {\mu }}_{{ {\scriptscriptstyle{\mathrm O}}_2 } } }}{{r_{{ {\scriptscriptstyle{\mathrm O}}_1 } { {\scriptscriptstyle{\mathrm O}}_2 } }^3 }}} [1 - K(r_{{ {\scriptscriptstyle{\mathrm O}}_1 } { {\scriptscriptstyle{\mathrm O}}_2 } } )]  - \alpha {\bf{E}}^{def},
\end{equation}
where the deformation field strength is
\[
{\bf{E}}^{def}  = \sum\limits_{j = 3,4}^{} {\frac{{b_1 e^{ - \rho _1 r_{{ {\scriptscriptstyle{\mathrm O}}_1 } j} } }}{{r_{{ {\scriptscriptstyle{\mathrm O}}_1 } j}^2 }}\left( {\rho _1  + \frac{1}{{r_{{ {\scriptscriptstyle{\mathrm O}}_1 } j} }}} \right)} {\bf{r}}_{{ {\scriptscriptstyle{\mathrm O}}_1 } j}  +
\]

\smallskip

\begin{figure}[!h]
\includegraphics[natwidth=1680, natheight=979, scale=0.141]{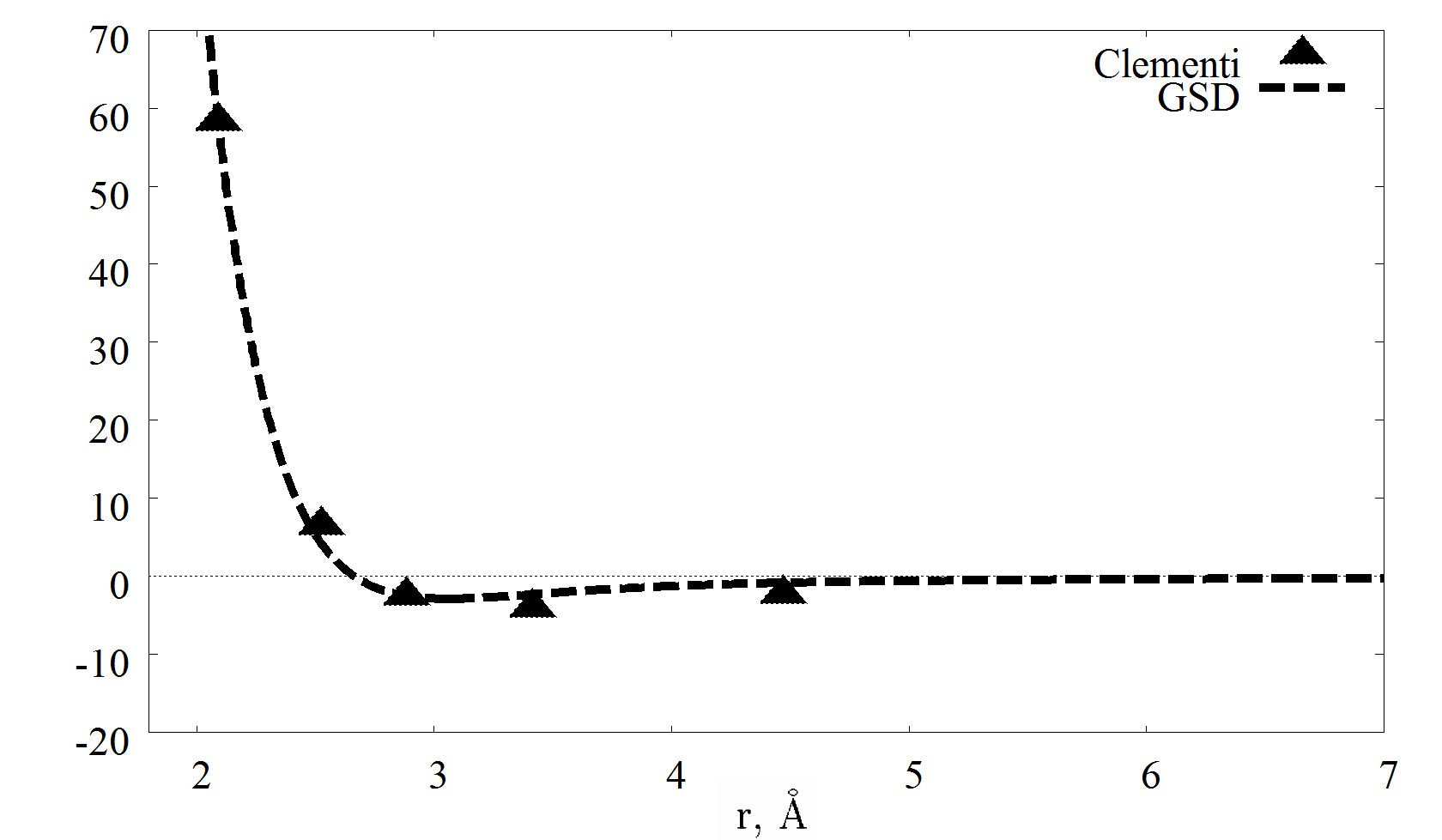}
\vskip 1mm\caption{Comparison of the interaction energies between two water molecules for the configuration depicted in Fig. 4
}
\end{figure}
\vskip-4mm
\begin{equation}
 + \frac{{b_2 e^{ - \rho _2 r_{{ {\scriptscriptstyle{\mathrm O}}_1 } { {\scriptscriptstyle{\mathrm O}}_2 } } } }}{{r_{{ {\scriptscriptstyle{\mathrm O}}_1 } { {\scriptscriptstyle{\mathrm O}}_2 } }^2 }}\left( {\rho _2  + \frac{1}{{ {\scriptscriptstyle{\mathrm O}}_1 } { {\scriptscriptstyle{\mathrm O}}_2 } }} \right){\bf{r}}_{{ {\scriptscriptstyle{\mathrm O}}_1 } { {\scriptscriptstyle{\mathrm O}}_2 } },
\end{equation}
and ${\bf{\hat T}}_{{ {\scriptscriptstyle{\mathrm O}}_1 } { {\scriptscriptstyle{\mathrm O}}_2 } }  = {\bf{\hat I}} - \frac{{3{\bf{r}}_{{ {\scriptscriptstyle{\mathrm O}}_1 } { {\scriptscriptstyle{\mathrm O}}_2 } }  \otimes {\bf{r}}_{{ {\scriptscriptstyle{\mathrm O}}_1 } { {\scriptscriptstyle{\mathrm O}}_2 } } }}{{r_{{ {\scriptscriptstyle{\mathrm O}}_1 } { {\scriptscriptstyle{\mathrm O}}_2 } }^2 }}$ is the tensor of dipole--dipole interaction. The deformation field strength is a
gradient of the repulsion force potential, ${{\bf{E}}^{def}  =  - {\boldsymbol{\nabla}} \Phi _{{\rm{II}}}}$.

Hence, the dipole moments ${\boldsymbol {\mu }}_{ {\scriptscriptstyle{\mathrm O}}_1 }$ and ${\boldsymbol {\mu }}_{ {\scriptscriptstyle{\mathrm O}}_2}$ of oxygens are functions of the parameters ($b_1$, $\rho _1$) and ($b_2$, $\rho _2$), respectively. The values of parameters $b_1$ and $\rho _1$ were
determined above (see formula (18)). To find the parameters $b_2$ and $\rho _2$, let us calculate, using formula (8),
the energy of interaction between two molecules in the
configuration depicted in Fig. 4.

For this configuration, five numerical values for the
interaction energy at various distances between oxygen atoms were obtained in work \cite{Kistenmacher} on the basis of quantum-chemical calculations.
The $b_2$- and $\rho _2$-values were so determined that the calculated curve optimally reproduced the positions of those points. The corresponding values obtained are
\[
{b_2=42129.1}, \, \, {\rho _2=2.59}.
\]

A comparison between the energies calculated by formula (8) and reported in work \cite{Kistenmacher} is shown in Fig. 5. Below, when calculating the interaction energy for two water molecules, just those $b_2$- and $\rho _2$-values were used.

In the MPM, the parameters $b_2$ and $\rho _2$ were determined using the interaction energy $\Phi _{(0)}  =  - 8.47$ between water molecules in a dimer at the distance  $r_{{ {\scriptscriptstyle{\mathrm O}}_1 } { {\scriptscriptstyle{\mathrm O}}_2 } }  = 2.96$~\AA \, between oxygen atoms (see work \cite{AntonchenkoDavydov}).

\subsection{ Dipole moment of an isolated water
molecule }

The electric dipole moment of an isolated water molecule
is determined as a sum of two antiparallel dipole-moment
vectors, ${\boldsymbol {\mu }} = {\boldsymbol {\mu }}_{\scriptscriptstyle{\mathrm H}}  + {\boldsymbol {\mu }}_{\scriptscriptstyle{\mathrm O}}$. The dipole moment ${\boldsymbol {\mu }}_{\scriptscriptstyle{\mathrm H}}$ is defined by the spatial distribution of the centers of negative oxygen and positive hydrogen charges, ${\boldsymbol {\mu }}_{\scriptscriptstyle{\mathrm H}}  = q_{\scriptscriptstyle{\mathrm H}} ({\bf{r}}_1  + {\bf{r}}_2 )$. The absolute value of dipole moment ${\boldsymbol {\mu }}_{\scriptscriptstyle{\mathrm H}}$ is $\mu _{\scriptscriptstyle{\mathrm H}}  = 2 q_{\scriptscriptstyle{\mathrm H}} r_{\scriptscriptstyle{\mathrm{OH}}} \cos ({\textstyle{1 \over 2}}\theta ) = 5.6281$~D. The dipole moment ${\boldsymbol {\mu }}_{\scriptscriptstyle{\mathrm O}}$ of an oxygen atom emerges owing to the polarization of the electron shell of an oxygen anion by the electric
fields of hydrogens in the water molecule.
According to the results of work \cite{StillingerDavid}, it equals
\vskip-4mm \[
{\boldsymbol {\mu }}_{\scriptscriptstyle{\mathrm O}}  =  - \alpha  q_{\scriptscriptstyle{\mathrm H}} \left( {\frac{{{\bf{r}}_1 }}{{r_1^3 }}\left[ {1 - K(r_1 )} \right] + \frac{{{\bf{r}}_2 }}{{r_2^3 }}\left[ {1 - K(r_2 )} \right]} \right) .
\]
\vskip-1mm
\noindent{It is easy to calculate that $\mu _{\scriptscriptstyle{\mathrm O}}  =  - 3.7752$~D. Therefore,
the magnitude of the dipole moment  ${\boldsymbol {\mu }}$ is   $\mu  = \mu_{\scriptscriptstyle{\mathrm H}}  + \mu_{\scriptscriptstyle{\mathrm O}}  = 1.8528$~D. This value completely agrees with the absolute
value of dipole moment of an isolated water molecule.}

\subsection{Influence of a neighbor molecule on the
dipole moment of a water molecule }

The modification of the dipole moment under the influence of a neighbor molecule is one of the simplest manifestations of many-particle effects in the system.  To estimate the influence of the second molecule, let us calculate the ratio ${\mu_{\scriptscriptstyle{\mathrm O}}^{(12)} } / {\mu_{\scriptscriptstyle{\mathrm O}} }$
between the dipole moment of an oxygen atom calculated in the pair approximation to the dipole moment of an isolated molecule.
The dependence of the ratio ${\mu_{\scriptscriptstyle{\mathrm O}}^{(12)} } / {\mu_{\scriptscriptstyle{\mathrm O}} }$
on the distance between the oxygen atoms of two neighbor molecules is exhibited in Fig. 6.

\section{Conclusions}

This work was devoted to the generalization of the
known Stillinger–David potential $\Phi_{\mathrm{SD}}$, which is widely
used for the description of the intermolecular interaction in water.
Two versions were considered: the modified Stillinger–David potential $\Phi _{\mathrm{MSD}}$ and the generalized Stillinger–David potential $\Phi _{\mathrm{GSD}}$. The modified potential $\Phi _{\mathrm{MSD}}$ improves the behavior of the potential $\Phi_{\mathrm{SD}}$ at rather large distances between water molecules. The generalized potential $\Phi _{\mathrm{GSD}}$ takes additionally the polarization effects into account more adequately in comparison with the initial potential $\Phi_{\mathrm{SD}}$. Moreover, the fact that the repulsion part of the interaction between oxygen atoms affects their polarization \cite{Zhyganiuk,ZhyganiukProceeding} was taken into account.

\begin{figure}[t]
\includegraphics[natwidth=1680, natheight=979, scale=0.128]{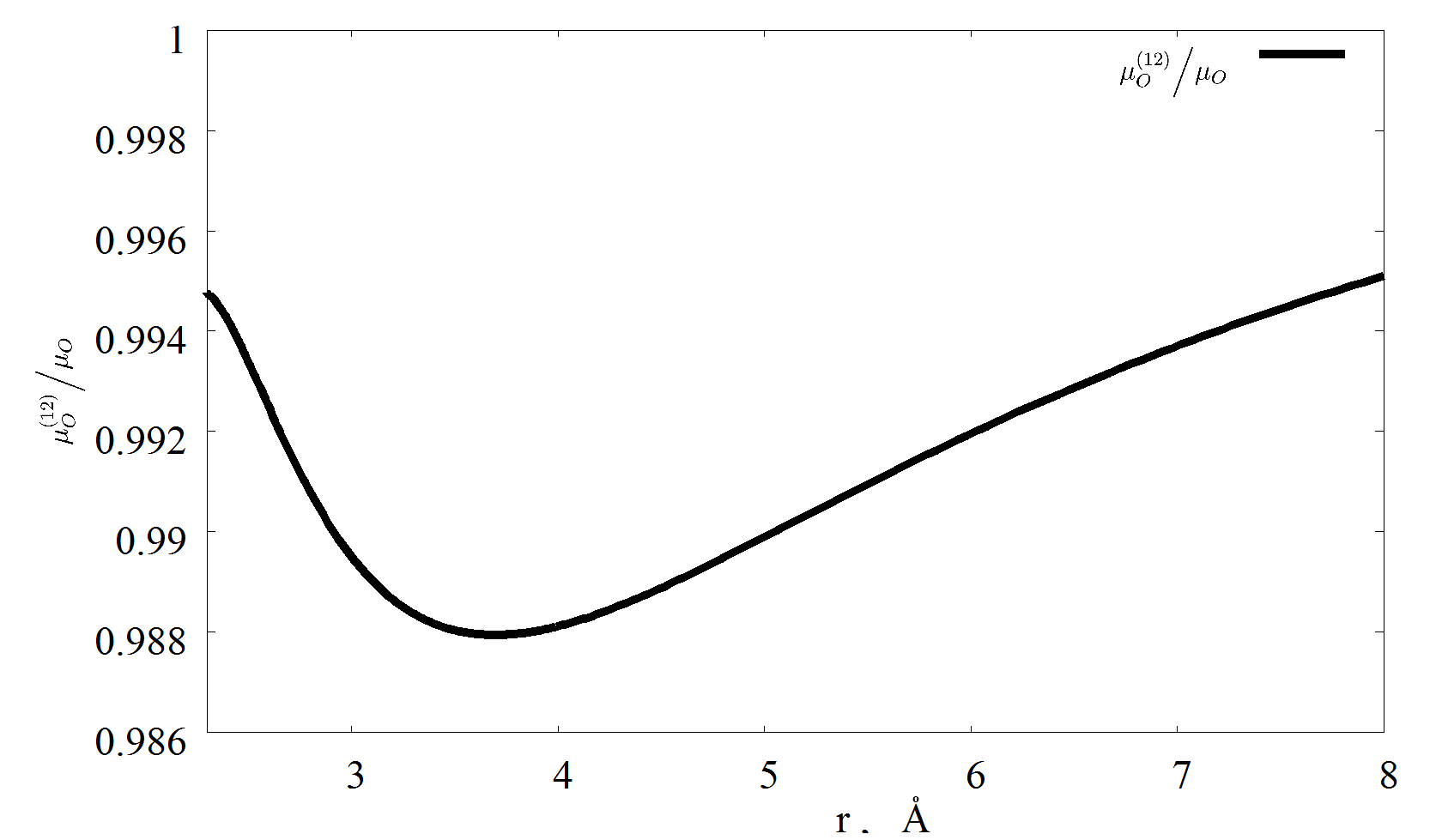}
\vskip-4mm\caption{Dependence of the ratio
${\mu_{\scriptscriptstyle{\mathrm O}}^{(12)} } / {\mu_{\scriptscriptstyle{\mathrm O}} }$ on the distance between
oxygen atoms in two water molecules
}
\end{figure}

The values obtained for the interaction energy between water molecules with the use of the generalized Stillinger–David potential and in quantum-chemical calculations \cite{Kistenmacher} coincide within the whole range of inter-molecular distances. The results of detailed calculations
of the ground state energy and the dimer vibrational frequencies on the basis of $\Phi _{\mathrm{MSD}}$ and $\Phi _{\mathrm{GSD}}$ potentials will
be presented in a separate work.

\medskip

The author expresses his sincere gratitude Professor M.P. Malomuzh for his permanent attention and numerous long discussions concerning all issues dealt with in the work. I am sincerely thankful  Academician of the National Academy of Sciences of Ukraine
L.A. Bulavin for his support of the work at every stage
of its performance, as well as for the opportunity to discuss the results obtained at the seminar of the Chair of
Molecular Physics at the Faculty of Physics of the Taras
Shevchenko National University of Kyiv.

\vskip 5mm
\begin{flushright}
{\footnotesize Received 29.12.10.} \\
{\footnotesize Translated from Ukrainian by O.I.~Voitenko}
\end{flushright}

\vskip -5mm
\rezume{%
УЗАГАЛЬНЕНИЙ ПОТЕНЦІАЛ СТІЛІНДЖЕРА І ДЕВІДА}{І.В. Жиганюк}{У роботі запропоновано вдосконалений поляризаційний потенціал Стілінджера і Девіда для міжмолекулярної взаємодії у воді. Сформульовано чіткий алгоритм визначення функції, яка описує взаємодію оксиген--гідроген в молекулі води. Розроблено новий підхід до моделювання функції, що екранує заряд--дипольну взаємодію на малих відстанях. Для правильного опису асимптотичної поведінки міжмолекулярного потенціалу на достатньо великих відстанях потенціал Стілінджера і Девіда завершено взаємодією між дипольними моментами оксигенів. Крім того, поляризаційна складова потенціалу Стілінджера і Девіда доповнена доданком, що описує деформацію електронних оболонок оксигенів. Узагальнення потенціалу Стілінджера і Девіда дозволяє успішно відтворити всі основні результати квантово-хімічних розрахунків енергій взаємодії двох молекул води, отриманих Клементі. Вивчено поведінку дипольного моменту молекули води як функції міжмолекулярної відстані та отримано оцінку незвідних двохчастинкових ефектів у воді.
}


\begin{thebibliography}{22}

\bibitem {BulavinKarm} L.A.~Bulavin, T.V.~Karmazina, V.V.~Klepko, and
V.I.~Slisenko , Neutron Spectroscopy of Condensed Media (Akademperiodika, Kyiv, 2005) (in Ukrainian).

\bibitem {BulavinFisen} L.A.~Bulavin, A.I.~Fisenko, and N.P.~Malomuzh, Chem.~Phys. Lett. \textbf{453}, 183 (2008).

\bibitem {AdamenkoBulavin} I.I.~Adamenko and L.A.~Bulavin, Physics of Liquids and
Liquid Systems (ASMI, Kyiv, 2006) (in Ukrainian).

\bibitem {AntonchenkoDavydov} V.Ya.~Antonchenko, A.S.~Davydov, and V.V.~Il’in, Fundamentals of Physics of Water (Naukova Dumka, Kyiv, 1991) (in Russian).

\bibitem {AntonchenkoIlyin} V.Ya.~Antonchenko, V.V.~Ilyin, N.N.~Makovskii, and
S.A.~Polesya, Dokl. Akad. Nauk UkrSSR, No. 8, 41 (1985).

\bibitem {Antonchenko} V.Ya.~Antonchenko, Physics of Water (Naukova Dumka, Kyiv, 1986) (in Russian).

\bibitem {Bernal} J.D.~Bernal and R.H.~Fowler, J.~Chem. Phys. \textbf{1}, 515 (1933).

\bibitem {RahmanStillinger} A.~Rahman and F.H.~Stillinger, J.~Chem. Phys. \textbf{55}, 3336 (1971).

\bibitem {Jorgensen} W.L.~Jorgensen, J.~Chandrasekhar, and J.D.~Madura, J.~Chem. Phys. \textbf{79}, 926 (1983).

\bibitem {Kistenmacher} H.~Kistenmacher, G.C.~Lie, H.~Popkie, and E.~Clementi, J.~Chem. Phys. \textbf{61}, 546 (1974).

\bibitem {Poltev} V.I.~Poltev, T.A.~Grokhlina, and G.G.~Malenkov, J.~Biomolec.~Struct.~Dynam. \textbf{2}, 413 (1984).

\bibitem {StillingerDavid} F.H.~Stillinger and C.W.~David, J.~Chem. Phys. \textbf{69}, 1473 (1978).

\bibitem {StillDav} F.H.~Stillinger and C.W.~David, J.~Chem. Phys. \textbf{73}, 3384 (1980).

\bibitem {Lokotosh} T.V.~Lokotosh, Zh. Fiz. Khim.  \textbf{67}, 210 (1993).

\bibitem {Zabrodskii} V.G.~Zabrodskii and T.V.~Lokotosh, Ukr. Fiz. Zh. ~\textbf{38}, 1714 (1993).


\bibitem {LokotoshMalomuzh} T.V.~Lokotosh and N.P.~Malomuzh, Atti Acc. Pelor.  Pericol. Classe I di Sci. Fis. Mat. Nat., \textbf{72}, 2 (1994).

\bibitem {ZabrodskiiLokotosh} V.G.~Zabrodskii and T.V.~Lokotosh, Ukr. Fiz. Zh. \textbf{40}, 693 (1994).

\bibitem {Eisenberg} D.~Eisenberg and W.~Kauzmann, The Structure and Properties of Water (Oxford Univ. Press, New York, 1969).

\bibitem {Smith} D.F.~Smith, jr. and J.~Overend, Spectrochim. Acta A \textbf{28}, 471 (1972).


\bibitem {Tikhonov} A.N.~Tikhonov, Dokl. Akad. Nauk SSSR \textbf{163}, 591 (1965).

\bibitem {Zhyganiuk} I.V.~Zhyganiuk, Dopov. Nat. Akad. Nauk Ukr. No. 8, 77
(2009).


\bibitem {ZhyganiukProceeding} I.V.~Zhyganiuk, Dopov. Nat. Akad. Nauk Ukr. No. 11,
72 (2009).

\end{thebibliography}
\end{document}